\documentclass[prl,twocolumn,showpacs,showkeys,floatfix]{revtex4}
\usepackage{latexsym,amssymb,amsmath,amsfonts}
\usepackage{graphicx}
\usepackage{bm}
\usepackage{setspace}
\newcommand{\beq}{\begin{equation}}
\newcommand{\eeq}{\end{equation}}
\newcommand{\bc}{\begin{center}}
\newcommand{\ec}{\end{center}}
\newcommand{\eeqa}{\end{eqnarray}}
\newcommand{\beqa}{\begin{eqnarray}}
\newcommand{\no}{\noindent}

\newcommand{\ra}{\rightarrow}

\newcommand{\ga}{\gamma}
\newcommand{\Ga}{\Gamma}
\newcommand{\de}{\delta}
\newcommand{\De}{\Delta}
\newcommand{\ep}{\epsilon}

\newcommand{\si}{\sigma}

\newcommand{\ph}{\phi}

\newcommand{\ps}{\psi}

\newcommand{\om}{\omega}
\newcommand{\ed}{\end{document} }
\begin{document}

\title{Spin flip of electron in static electric fields}
\author{Richard T. Hammond}

\email{rhammond@email.unc.edu }
\affiliation{Department of Physics\\
University of North Carolina at Chapel Hill\\
Chapel Hill, North Carolina and\\
Army Research Office\\
Research Triangle Park, North Carolina}

\date{\today}

\pacs{31.15.aj, 31.30.jc}

\keywords{Spin flip }

\begin{abstract} The effects on the spin state of an electron in a time independent electric field are examined. The probability of spin flipping is calculated, and other effects are studied using the minimally coupled Dirac equation.
\end{abstract}

\maketitle

\section{introduction}
There is a vast yet growing interest in spin states and their control. This attention comes from a number of sectors, from the quantum information world and the need to spin flip and control the spin, to a myriad number of application of spintronics. It is well-known how to flip a spin, and basic physics tells us the most obvious way is through the use of a magnetic field. This is because the interaction between a charged particle with spin and an external electromagnetic field is ${\bm \mu}\cdot \bm B$. But this is the non-relativistic limit. We know moving through a $\bm B$-field with velocity $v$ produces an electric field, so we expect electric field effects  to go like $v/c$ times the magnetic effect. 

In fact, it is well known spin can be changed without using a magnetic field.
Nuclear spin flips induced by an electric field have recently been observed,\cite{sugimoto} and electric field effects with a laser were demonstrated a while ago.\cite{mooradian} Electric fields are also known to cause spin flipping through the Rashba effect,\cite{rashba} and dynamic spin process such as the Kapitza-Dirac effect are well known.\cite{ahrens} Spins can also be flipped using non-electromagnetic means.\cite{hammondprd}

Although, at first glance, one expects the magnetic field to be used in the control of spin,  there are other reasons for considering electric effects. In small devices, electric fields are usually easier to make, both from size considerations of power usage, so it is worthwhile to investigate a little more carefully the effect of the electric field on spin. In addition, if there is a current present, then there is an electric field present as well.

The main objective of this paper is to examine and assess the effect of a known external electric field on the spin state of an electron. In the applications given above, it is impossible to isolate the  effects of an electric field. This is because many of those effects are in various shaped bulk materials with an electric field of unknown functional form, some of which have current density or accompanying magnetic fields. Under these
conditions it is impossible to really know the effect of the electric field alone. In addition, may times the properties of the material is modeled by a Hamiltonian that is assumed to hold, at least at some approximation. In this paper, we treat the problem of an electron in an electric field exactly (although we stop at first order perturbation), the interaction of which comes directly from the Dirac equation. Other than using first order theory, no approximations are made, and we can be confident we are seing the effects of the electric alone.

Before the effect of the electric field is considered, there is an important conceptual issue that should be addressed. Let us consider an electron in the non-relativistic regime. Then we say it has either spin up or spin down (or spin right or spin left, etc.), reflecting the two allowed states of spin for the electron. But spin up or down are meaningless concepts without a fiduciary field by which to measure them. So we place the electron in an external magnetic field and now the concept of spin up and down makes sense.

In particular, we consider flipping the spin. We should not think of flipping the spin of a free particle for the reasons discussed, so the initial and final states are not those of a free particle, they are those of an electron in a magnetic field, which is assumed to be constant. It has been shown that, assuming the final and initial states are those of the eigentstates of an electron in a magnetic field, then the scattering matrix is\cite{hammondapl}
 
\beq\label{sfi}
S_{fi}=-i\frac{e}{\hbar c}\int d^4x \overline \psi_f A_\si\ga^\si \ps_i
\eeq

\no which, of course, looks the same as the free particle form, but it is important to note that in (\ref{sfi}) the initial and final states are those of a particle in a magnetic field. The above formula is given in cgs units, but below natural units ($\hbar=1=c$)
are used in general, although some formulas are given in cgs for clarity, including the final results.

\section{known solutions}
For convenience, this brief section will summarize known properties of an electron in a constant magnetic field $B$.\cite{huff} The energy is quantized in Landau levels with quantum number $n$
\beq\label{en}
E_n^2=m^2+p_z^2+2neB
\eeq
where $p_z$ is the $z$ component of the momentum, $e$ is the magnitude of the charge of the electron, and $n=0,1,2...$,
\no and the wavefunctions are given by

\beq
\psi_n=C_fe^{-i(E_nt-p_x^nx-p_z^nz)}e^{-\xi^2/2}u_n
\eeq

\no where the momentum terms are eigenvalues, 
and the dimensionless coordinate is

\beq
\xi=\sqrt{eB}y -\frac{p_x}{\sqrt{eB}}
.\eeq

In terms of the energy quantum number $n$ the solutions are given by
for spin up,
\beq
u_n=\left(
\begin{array}{cccc}
h_{n-1} &\\
0 &\\
p_z^nh_{n-1}/(E_n+m) &\\
-\sqrt{2neB}h_n/(E_n+m) &\\
\end{array}\right)
\eeq

\no and for spin down,

\beq
u_n=\left(
\begin{array}{cccc}
0 &\\
h_n &\\
-\sqrt{2neB}h_{n-1}/(E_n+m)  &\\
-p_z^nh_n/(E_n+m) &\\
\end{array}\right)
\eeq

\no where $u_n=u_n(\xi)$ 
and $h_n=h_n(\xi)$ are the orthonormal Hermite polynomials, 
 $h_n=N_nH_n$, and $H_n$ are the Hermite polynomials, $N_n=1/\sqrt{2^nn!\sqrt{\pi}}$, and by definition the Hermite polynomial with a negative subscript is zero.

The solution is normalized to a two-dimensional box in the $x$ and $z$ directions, and the normalization constants are

\beq
C=\sqrt{\frac{(E_n+m)}{2L_\xi L_xL_zE_n}},
\ \ \ \ \ \ L_\xi\equiv \sqrt{\frac{\hbar c}{eB}}
\eeq
where $L_\xi$ is given in cgs.
It is useful to note the solutions have a two-fold degeneracy.

\section{scattering amplitude}

We are now in a position to describe all the effects of an electromagnetic perturbation, $A_\si$ on an electron trapped in a magnetic field. Using the above in (\ref{sfi}) we have

\beqa\label{sfi2}
S_{fi}=
-\frac{iC_fC_i}{L}\int d^4x 
e^{i(E_ft-p_x^fx-p_z^fz)}e^{-\xi^2}\nonumber \\
M_{fi}e^{-i(E_it-p_x^ix-p_z^iz)}
\eeqa
(to avoid double subscripting, here and below, when $f$ or $i$ appears as a subscript or superscript, it really stands for $n_f$ or $n_i$, the final or initial quantum number)
where
$L\equiv \hbar\om/eE$ and
the matrix element is

\beq
M_{fi}=u_f^\dag\ga^0\ga^\si A_\si u_i
,\eeq
which becomes,
\begin{widetext}
\beqa\label{m}
M_{fi}=\ \ \ \ \ \ \ \ \ \ \ \ \ \ 
A_0 \left(\frac{\sqrt{2}
   \sqrt{B n_f e} h_f h_i
   p_z^i}{(E_f+m)(E_i+m)}
   -\frac{\sqrt{2}
   \sqrt{B n_i e} h_{f-1} h_{i-1}
   p_z^f}{(E_f+m)(E_i+m)}\right)
+A_1
   \left(\frac{h_{f-1}h_i
   p_z^f}{(E_f+m)}-\frac{h_{f-1}h_i
   p_z^i}{(E_i+m)}\right)\\ \nonumber
-iA_2
   \left(\frac{h_{f-1}h_i
   p_z^f}{(E_f+m)}-\frac{h_{f-1}h_i
   p_z^i}{(E_i+m)}\right)
+A_3 \left(\frac{\sqrt{2} \sqrt{B
   n_f e} h_f
   h_i}{(E_f+m)}-\frac{\sqrt{2} \sqrt{B n_i e} h_{f-1}
   h_{i-1}}{(E_i+m)}\right)\\ \nonumber
.\eeqa   
\end{widetext}

This last term has a wealth of information. For example, for a static electric field the only contributions come from the $A_0$ term, whereas magnetic effects are described by the other terms.  If one compares the $A_0$ term to the $A_3$ term we find the ratio is goes like $pc/mc^2\sim v/c$. This is expected, and we also note the $A_0$ term vanishes unless the electron is moving, (we know an observer moving through and electric field sees a magnetic field $\bm {v}\times{\bm B}/c)$.

Another interesting situation arises from the fact the energy levels are determined by two parameters, the momentum and the magnetic interaction energy. Thus, the electron can suffer a spin flip  even though the total energy of the system remains constant. This can be seen by looking at  the energy levels and noting, from (\ref{en}),
$E\De E = p\De p+ \De n eB$, so that the total energy can remain constant while the electron gains (loses) magnetic energy as the electron loses (gains) kinetic energy.

\subsection{Rabi formula}
Here we derive a known result. There are several reason to do this, one is to verify the formalism, but it also puts the result on a firm theoretical formulation. Also, we will have at our disposal the relativistic form of this eeuation. Finally it will display the kinds of calculations needed below.

In this case we write $A_1= (E/\om)f$ where $f= \sin(kz-\om t)$. Several definitions are useful to simplify things, and the dimensionless coordinate $\xi$ becomes
\beq
y=L_\xi(\xi-\ph),\ \ \ \ph=\frac{p_xc}{\sqrt{eB\hbar c}},
\eeq
so (\ref{sfi2}) yields

\begin{widetext}
\beq
S_{fi}=-\frac {LC_iC_f}{2}\int d^4xe^{-i(E_i-E_f)t}e^{i(p_f^x-p_i^x)x}e^{i(p_f^z-p_i^z)z}fe^{-\xi^2}(e^{i(kz-\om t)}-e^{-i(kz-\om t)})
{\mbox \cal M}_{fi}
.\eeq
\end{widetext}

\no The integral over $z$ yields $L_z^\pm=2\pi\de(p_f^z-p_i^z\pm k)$ and similarly $L_x$ (with no $k$).
The time integral, $L_t$, leads to
 $|L_t|^2=(\frac{\sin (\om-\om_0) t/2}{(\om-\om_0)/2})^2$ where $\om_0=E_f-E_i$. 
 \no The matrix element boils down to

\beq
\left(
\frac{h_f^2p_z^f}{E_f+m}-\frac{h_f^2 p_z^i}{E_i+m}
\right)
,\eeq
and using the orthogonality of the Hermite polynomials we may perform the final integral. Assuming the electron may either flip from down to up or vice versa, we have
 
 \beq
S_{fi}=\frac{C_fC_iL_yL_xL_tL_\xi}{2L}\frac{\hbar k}{mc}
\eeq

The spin flip probability is the integral of $|S_{fi}|^2$ times the density of states. Details were given elsewhere\cite{hammondapl} and the result is, in the low velocity limit

\beq
S_{fi}=\kappa\left(\frac{\sin (\om-\om_0) t/2}{(\om-\om_0)/2}\right)^2
\eeq
where $\kappa =E^2\mu^2/\hbar^2$ and $\mu=e\hbar/2mc$. This is the well-known Rabi formula.

\section{Time independent electric field}
Now we would like to consider the effect of an external electric field. Since a time dependent electric field creates a magnetic field,  only ``pure electric'' effects can be seen from a static field, which we assume is imposed externally. The only other requirement is that the field obey Maxwell's equations, for otherwise we would end up with unrealistic effects. A relatively simple electric field is one is which the potential is given by
$A_0=(E/k)\sin kz e^{-ky}$ where $E$ is the electric field strength and $k$ is a parameter describing the spatial variation of the field. With this we may perform the integrations is (\ref{sfi2}) and obtain the transition matrix. A few details will discussed.

This time the time integral, $L_t$, is of the form $\int e^{-i(E_i-E_f)t}dt$. Integrating from minus to plus infinity gives a delta function, showing $E_i=E_f$ (recall from above that this can induce a spin flip through a loss or gain in momentum). In order to perform the integral over some finite time, we assume $E_i-E_f=\ep$ and take the limit as $\ep\ra0$. The square of this gives $4t^2$. The $x,z$ integrations give $L_x, L_z$, which are conservation of momentum and in particular $\de(p_z^f-p_z^i\pm\hbar k)$.  That leaves the $y$ integral, which may be accomplished using the orthogonality of the Hermite polynomials. The result is

\beqa
S_{fi}= -i\frac{C_iC_f}{L}L_tL_xL_zL_\xi\Ga\sum_{n=0}^{n_f}\\ \nonumber
\left(\begin{array}{cc}
	n_f&\\
	n
\end{array}\!\!\!\! \right)
\left(\begin{array}{cc}
	n_i&\\
	n
\end{array}\!\!\!\! \right)
\sqrt{n_f}(-kL_\xi)^{n_f+n_i-2n}p_z^i -R
\eeqa

\no where the terms with parentheses are the binomial coefficients, 

\beq
\Ga=\frac{\sqrt{2eB}}{(E_f+m)(E_i+m)}e^{kL_\xi \ph}e^{(kL_\xi/2)^2}
\eeq
and $R$ is the same as the term preceding it with $n_f$ replaced by $n_{f-1}$,
$n_i$ replaced by $n_{i-1}$, and $p_z^i$ replaced by $p_z^f$.

In order to find the transition probability $|S_{fi}|^2$ is integrated over the density of states as before. Let us simplify things a bit by assuming the momentum in the $x$ direction is small. This means we can set $\ph$ to zero (the momentum would have to be relativistic for this term to be important).
Now let us consider $k L_\xi$. For a $B$ field of 100 G, this would be on the order of $2\times 10^{-5}k$ ($k$ in cm) and will assumed to be small. This is valid unless the electric field varies appreciably over the scale of a tenth of a micron. However if such fields are imposed, then this term, being exponential, will have a dramatic effect on the spin flip probability.

Let us also assume  $n_i=0$. The spin flip probability $\cal W$ becomes

\beq
{\mbox{\cal W}}= \frac{B(p_z^f)^2E^2e^3}{m^4k^2}t^2
\eeq
but we must impose the conditions of the delta functions: From the momentum delta function we have
$p_f=p_i-k$ and from energy we have $p_f^2=p_i^2-2eB$.

\subsection{electric cooling}

As discussed above, it is possible for the scattering amplitude to be non-vanishing with the total energy remaining constant. In this case, if the spin flips from a higher to lower state, the energy is given to kinetic energy of the electron, as the delta functions dictate. On the other hand, if the spin flips to a higher energy state, the electron and loses that much kinetic energy, cooling the sample. The amount of cooling can be controlled by the amplitude of the applied electric and magnetic fields. 
Laser cooling with spin has been discussed for condensed matter systems.\cite{togan}

As an example, for the $z$ component of the momentum, setting $p^f$ to zero we find $p^i=\mu B/2mc^2$, which we expect on simple energy considerations.

\section{summary}
 The main goal has been achieved, to show that an electron moving through a static electric field can suffer spin flips. The transition amplitude was calculated using the Dirac equation, and the spin flip probability for an electron was calculated from that. The Rabi formula was derived from the formalism, although a large number of relativistic effects may be gleaned from (\ref{m}). In the present case, we examined the effect of a static electric field.

\ed